\documentclass[twocolumn,prl,a4paper,superscriptaddress,showpacs]{revtex4}

\usepackage[english]{babel}
\usepackage{hyperref}
\usepackage{graphicx}
\usepackage{amsmath}
\usepackage{mathrsfs}

\begin{document}

\title{SQUID Detection of Quantized Mechanical Motion}

\author{Stefano Pugnetti}

\affiliation{NEST-CNR-INFM and Scuola Normale Superiore, Piazza dei
  Cavalieri 7, I-56126 Pisa, Italy}  

\author{Yaroslav M.~Blanter}

\affiliation{Kavli Institute of Nanoscience, Delft University of
  Technology, Lorentzweg 1, 2628 CJ Delft, The Netherlands }  

\author{Rosario Fazio}

\affiliation{NEST-CNR-INFM and Scuola Normale Superiore, Piazza dei
  Cavalieri 7, I-56126 Pisa, Italy}  

\begin{abstract}
We predict that quantized mechanical motion can be detected by
embedding a mechanical resonator into a quantum SQUID. If the system
is tuned to the regime when a plasma frequency of the SQUID matches
the resonator frequency, the doubly-degenerate quantum level of the
system is split by the coupling between the SQUID and the
resonator. Observation of an avoided crossing as the function of
external flux would be an unambiguous evidence of quantum nature of
mechanical motion. We also investigate the conditions maximizing the
level splitting. 
\end{abstract}


\maketitle

The interest in nanoelectromechanical systems (NEMS) has been growing
rapidly~\cite{Cleland,Blencowe04} because of their wide range of
potential technological applications in detection and sensing, and
their importance in testing fundamentals of quantum theory. The
possibility to observe quantum mechanical motion of an oscillator has
important implications in the understanding to which extent
macroscopic objects obey the laws of quantum
mechanics~\cite{Leggett02}. NEMS can play also an important role in
quantum computation where they have been proposed as
qubits~\cite{Savel'ev06}, memory
elements~\cite{Cleland04,Pritchett}, and quantum
buses~\cite{Cleland04,Zou04}. Coupling of nanomechanical
oscillators to a qubit has been also thoroughly investigated in the
literature and many schemes of this kind have been proposed with
Cooper-pair
boxes~\cite{Armour02,Martin04,Rabl04,Wei06,Armour08,Hauss08},
Josephson junctions (phase
qubits)~\cite{Cleland04,Trees07}, quantum point
contacts~\cite{Ruskov05}, and quantum dots\cite{Liao08}. Recently the
dispersive coupling of a NEMS to a Cooper-pair box has been
realized~\cite{LaHaye09}.  

Coupling of nanomechanical oscillators to SQUIDs has been recently
intensively
investigated~\cite{Zhou06,Buks06,Xue07,Wang08,Zhang09,Pugnetti09b}.
The high sensitivity of SQUIDs to tiny changes in the magnetic flux
has suggested that the position of a nanomechanical resonator could be
monitored by integrating the oscillator into the superconducting loop
of a dc SQUID; indeed the transport properties of this superconducting
circuit in presence of a uniform magnetic field depend on the position
of the oscillator, since this position modifies the total area
threaded by the flux. Recently this scheme has been demonstrated for
the detection of the thermal motion of a mechanical resonator in the
classical regime~\cite{Etaki08}. Due to the high degree of control
achieved on quantum SQUIDs, coupling nanomechanical
resonators to SQUIDs is a promising scheme for observing quantum
effects in the motion of these oscillators as well.  

In this Article, we develop a protocol of detecting quantized
mechanical oscillations with a SQUID coupled to a mechanical
resonator. We show that the signature of this quantized motion is a
splitting of an energy level associated with the SQUID due to the
coupling to a resonator. In the case of resonant coupling, this
splitting can be detected by standard techniques developed for flux
qubits~\cite{Mooij99}. We stress that achieving this
regime is not straightforward as the typical frequencies of the SQUID (of the
order of few GHz) and the oscillator (MHz range) do not
naturally match to allow for a resonant behavior. In this
Article we show that this is however possible by tuning the external
magnetic field and the bias current with available experimental
setups. 

\begin{figure}[!tbh]
 \includegraphics[width=\columnwidth]{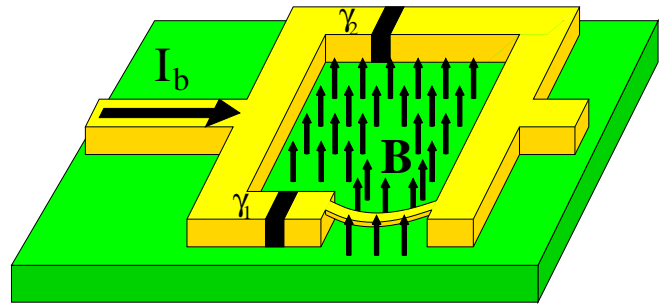}
 \caption{\label{fig:device}A sketch of the device we study: one arm
 of a dc SQUID is free to oscillate in the plane of the SQUID
 itself. A uniform magnetic field $B$ orthogonal to the SQUID plane is
 present and a dc bias current $I_b$ flows through the device. The two
 Josephson junctions, whose phase drops are respectively $\gamma_1$
 and $\gamma_2$, are taken to be identical.}  
\end{figure} 

The system we consider is schematically drawn in
Fig.~\ref{fig:device}. A dc SQUID is made of a superconducting loop of
total area $A$ with an arm of length $l$ that can oscillate freely in
the plane of the loop. For simplicity we assume that only a single
mode of oscillation with the frequency $\omega$ can be excited. This
mode is described by the dynamical variable $X$ representing the shift
of the center-of-mass of the resonator (with the mass $m$) with
respect to its rest position. The quantum effects related to
mechanical motion of this oscillator appear at the scale of the
amplitude of 
zero-point motion $X_0=\sqrt{\hbar/2m\omega}$. The SQUID further
comprises two Josephson junctions of equal critical currents $I_c$ and
shunting capacitances $C$; the typical energy scales related to the
physics of the junctions are the Josephson energy $E_J=\hbar I_c/2e$
and the charging energy $E_c=(2e)^2/2C \ll E_J$, whose magnitude
depends on the geometry of the junctions. The typical time scale for
the dynamics of the junctions is set by the inverse plasma frequency
$\omega_\mathrm{pl}=\sqrt{2E_JE_c}/\hbar$. The dynamics of the SQUID
is described by the two gauge-invariant phase drops $\gamma_1$ and
$\gamma_2$ across the two junctions. 

The coupling of the SQUID dynamics to mechanical motion is provided by
the position dependence of the magnetic flux threading the SQUID
loop. 
The two phases are constrained by the requirement that the
superconducting order parameter is single valued, 
\begin{equation}
 \label{eq:constraint}
 \gamma_1-\gamma_2=2\pi\left(\frac\Phi{\Phi_0}+n\right)\,,
\end{equation}
where $\Phi_0=h/2e$ is the flux quantum; 
the total flux $\Phi$ is the sum of three contributions. The first one
comes from the external bias $\Phi_e=BA \equiv\Phi_0\phi_e$, while the
second one $\Phi_m=BlX$ depends on the position of the mechanical
resonator and provides the coupling between the mechanical resonator
and the SQUID. If the circuit also has non-negligible self inductance
$L$, the third contribution to the total flux $\Phi$ comes from the
current circulating in the loop; if $I_1$ and $I_2$ are the currents
flowing through each junction, the self-induced flux reads
$L(I_1-I_2)/2$. This device has three degrees of freedom; if $L=0$,
then the constraint expressed by Eq.~(\ref{eq:constraint}) reduces the
number of degrees of freedom from three to two. In the following, we
assume that the dissipation effects are negligible. 

Quantum coherence in the motion of the mechanical resonator can be
detected via spectroscopic measurements on the quantum SQUID. Indeed,
the energy levels associated with the SQUID
degrees of freedom are shifted due to the coupling to the
oscillator. However, this shift is of the
second order in the coupling and for realistic experimental parameters
is very small. We employ therefore below a different scheme, which
provides the result which is of the first order in the coupling. One
first chooses an appropriate set of values for the externally
controllable quantities $I_b$ and $\Phi_e$ such that the system can be
trapped in a minimum in which one of the eigenfrequencies of the
electromagnetic modes is the same as the frequency of mechanical
oscillations (to be referred below as {\em degeneracy condition});
then one moves slightly away from this degeneracy condition by
changing the remaining external parameter. A plot of the energy levels
as functions of the external parameters should display an avoided level
crossing, which is a clear indication that the SQUID is coupled to a
coherent quantum system.  

It is useful to describe the system in terms of three dimensionless
variables $\gamma=(\gamma_1+\gamma_2)/2$, $\phi=\Phi/\Phi_0$ and $\xi
= BlX/\Phi_0$. The potential energy of the system reads  
\begin{equation}
\label{eq:U}
\begin{split}
  U=E_J&\left[
    -2\cos\gamma\cos(\pi\phi)-
    \frac{I_b}{I_c}\gamma-\pi\frac{I_b}{I_c}\xi+\frac{\xi^2}{2\mathscr{A}^2} 
+ \right.\\    
    &\left.\quad+\frac{2\pi}{\beta_L}(\phi-\xi-\phi_e)^2
  \right],
\end{split}
\end{equation}
where we have introduced the parameters
\begin{equation}
 \mathscr{A}=\sqrt{\frac{2 E_J}{\hbar \omega}}\frac{BlX_0}{\Phi_0}\,;
\  \beta_L = \frac{2LI_c}{\Phi_0} \ .
\end{equation}
The parameter $\mathscr{A}$ is
proportional to the flux threading the area swept by the mechanical
resonator and plays the role of
a coupling parameter, as we show below. Typical values of
$\mathscr{A}=4\times10^{-5}$ can be obtained assuming $I_c=1\mu$A,
$\omega=1$GHz, $l=1\mu$m, $X_0=10$fm and $B=0.1$T. 
When the temperature is low
enough for the system to reach the quantum regime, the coordinates
typically oscillate around a minimum of the potential corresponding to
the values $\bar\gamma$, $\bar\phi$ and $\bar\xi$. We can approximate
the dynamics as a three-dimensional
harmonic oscillator, with the energy being a quadratic form, 
\begin{equation}
\label{eq:cl_ham}
\begin{split}
  E_\mathrm{tot}=\sum_{i}&
    \frac{\hbar^2}{2E_c}\dot q_i^2 +\sum_{i,j}E_JV_{ij}q_i q_j \ , 
\end{split}
\end{equation}
where the coordinates $q_i$ read
\begin{equation}
\label{eq:first_coos}
q_1=\gamma-\bar\gamma; \ 
    q_2=\pi(\phi-\bar\phi); \ 
    q_3=\dfrac{1}{\Omega}\dfrac1{\sqrt{2}\mathscr{A}} (\xi-\bar\xi)\,,
\end{equation}
and
\begin{equation}
\label{eq:V}
 V=\left(
    \begin{array}{@{}ccc@{}}
      r & -s& 0\\
      -s & r+ \dfrac2{\pi\beta_L}& -\dfrac{2\sqrt2\mathscr{A}}{\beta_L}\Omega\\
      0 &-\dfrac{2\sqrt2\mathscr{A}}{\beta_L}\Omega&
      \left(1+\dfrac{4\pi\mathscr{A}^2}{\beta_L}\right)\Omega^2 
    \end{array}
  \right)\,.
\end{equation}
(We have introduced $r=\cos\bar\gamma\cos(\pi\bar\phi)$,
$s=\sin\bar\gamma\sin(\pi\bar\phi)$ and
$\Omega=\omega/\omega_\mathrm{pl}$). 
The coupling $V_{23}$ between the
mechanical resonator and the SQUID is proportional to the ratio
$\mathscr{A}/\beta_L$ (for devices with a low self-inductance this
behavior no longer holds, see below for discussion); the decoupled
regime can be obtained for either $B \to 0$ or $L \to \infty$. 
The coordinate $Q_1 =\gamma - \bar\gamma$, corresponding to one
of the electromagnetic degrees of freedom,
oscillates with the frequency $\omega_\mathrm{pl}\sqrt
{\cos\bar\gamma\cos(\pi\bar\phi)}$; 
therefore if a minimum $(\bar\gamma,\bar\phi,\bar\xi)$ is such that  
\begin{equation} 
\label{eq:deg_cond}
\cos\bar\gamma\cos(\pi\bar\phi)=\Omega^2\,,
\end{equation} 
the frequencies associated with the motion of average phase drop
$\gamma$ and oscillator motion coincide up to correction of second
order in $\mathscr{A}$; we show below that the equality is indeed
exact to all orders in the coupling. Since in the low-inductance limit
the coordinates $\phi$ and $\xi$ are not independent, it is convenient
to switch to the basis where the submatrix corresponding to $\xi$ and
$\phi$ is diagonalized. The analytic expression for the matrix in the
new basis is quite cumbersome; below we give this expression in
the limit in which the degeneracy condition (\ref{eq:deg_cond})
holds,  
\begin{equation}
\label{eq:Vp}
  V'=\left(
    \begin{array}{@{}ccc@{}}
      \Omega^2
      & -s\dfrac{\sqrt2\pi\mathscr{A}}{C}\Omega
      & \dfrac{s}{C}
      \\
      -s\dfrac{\sqrt2\pi\mathscr{A}}{C}\Omega
      &\Omega^2
      & 0
      \\
      \dfrac{s}{C}
      & 0 
      & \Omega^2+\dfrac{2C^2}{\pi\beta_L}
    \end{array} 
  \right)\ ,
\end{equation}
with $C=\sqrt{2\pi^2\mathscr{A}^2\Omega^2+1}$. The first and second row and
column correspond to the phase drop $\gamma$ and to the mechanical
degree of freedom, respectively. The parameters $I_b$ and $\Phi_e$
corresponding to the degeneracy condition are found if one solves the
equation set,
\begin{equation}
\label{eq:eqs}
 \left\{
 \begin{array}{l}
  I_b/I_c=2\sin\bar\gamma\cos(\pi\bar\phi)\\
  2\beta_L^{-1}(\bar\phi-\phi_e-\bar\xi)=-\cos\bar\gamma\sin(\pi\bar\phi)\\
  \bar\xi=4\pi\mathscr{A}^2\beta_L^{-1}
  \left(\bar\phi-\phi_e-\bar\xi+(\beta_L I_b)/(4I_c) \right)\\ 
  \cos\bar\gamma\cos(\pi\bar\phi)=\Omega^2 \,,
 \end{array}
 \right.
\end{equation}
where the first three lines are satisfied by stationary points of the
potential energy (\ref{eq:U}) and the fourth line is the degeneracy
condition. In a minimum one must have $V > 0$. The unknowns are the
coordinates $(\bar\gamma,\bar\phi,\bar\xi)$ of the minimum and
$\Phi_e$; we use $I_b$ to tune the system to the degeneracy
point\cite{foot1}.  

Next, we quantize the system. The Hamiltonian reads
\begin{equation}
\label{eq:ham}
 H=\sum_{i=1}^3 \hbar \omega_i a_i^\dagger a_i + \sum_{i\neq j}\dfrac 14
 \hbar
 \omega_\mathrm{pl}\dfrac{\omega_\mathrm{pl}}
 {\sqrt{\omega_i\omega_j}}V'_{ij}(a_i^\dagger+a_i)(a_j^\dagger+a_j)   
\end{equation} 
with $\omega_i=\omega_\mathrm{pl}\sqrt{V'_{ii}}$. When the condition 
(\ref{eq:deg_cond}) is fulfilled, the first excited
levels $|100\rangle$ and $|010\rangle$ are quasi-degenerate and the
Hamiltonian (\ref{eq:ham}) restricted to their subspace becomes  
\begin{equation}
 H=\left(
  \begin{array}{@{}cc@{}}
   \hbar\omega & \lambda\\
   \lambda &\hbar\omega
  \end{array}
\right)\,, \ \ \lambda=\dfrac 14 \hbar\omega_\mathrm{pl} \dfrac1\Omega
V'_{\xi\gamma}\,. 
\end{equation}

\begin{figure}[!tbh] 
 \includegraphics[width=\columnwidth]{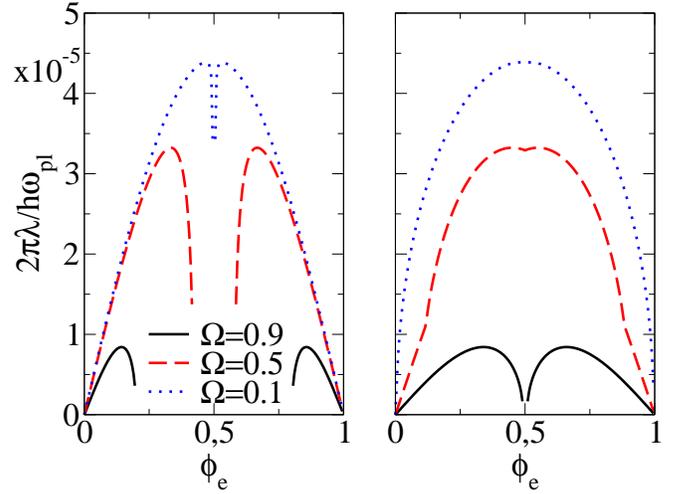}
 \caption{\label{fig:lambda_vs_phie} Dimensionless splitting
 $\lambda/\hbar\omega_\mathrm{pl}$ versus the external flux $\phi_e$
 for the coupling parameter $\mathscr{A}=4\times10^{-5}$. The
 self-inductance parameter $\beta_L$ is $10^{-4}$ for the left panel
 and $1$ for the right panel. Dotted, dashed, and solid lines
 correspond to $\Omega=\omega/\omega_\mathrm{pl} = 0.1, 0.5$ and
 $0.9$, respectively. The
 plots are periodic, the period being one flux quantum.}  
\end{figure}

Fig.~\ref{fig:lambda_vs_phie} shows the dependence of the dimensionless level
splitting $\lambda/\hbar\omega_\mathrm{pl}$ on the
dimensionless external flux $\phi_e$. 
The maximum level splitting is obtained for a
specific value of the external flux and depends very weakly on the
loop self-inductance; however the value of 
the flux at which the maximum is achieved, does; see below for further
discussion. The magnitude of the splitting is of the order of
$10^{-5}\hbar\omega_\mathrm{pl}$. The value of the ratio
$\Omega=\omega/\omega_\mathrm{pl}$ plays an important role: lower
values correspond to bigger maximum splittings and are thus
preferable. 

To enable the detection, the potential well formed at the chosen
minimum must be capable of containing quantum states. The number of
bound states can be estimated by the ratio between the energy
difference $\Delta U$ 
between the minimum and the closest saddle point and the energy level
separation, which is roughly $\hbar \omega$, 
\begin{equation}
\label{eq:splitting}
 \dfrac{\Delta U}{\hbar \omega}= \dfrac{\Delta
 U}{E_J}\dfrac1\Omega\sqrt{\dfrac{E_J}{2E_c}}\equiv\Delta
 u\sqrt{\dfrac{E_J}{2E_c}}\, .
\end{equation}

\begin{figure}[!tbh]
 \includegraphics[width=\columnwidth]{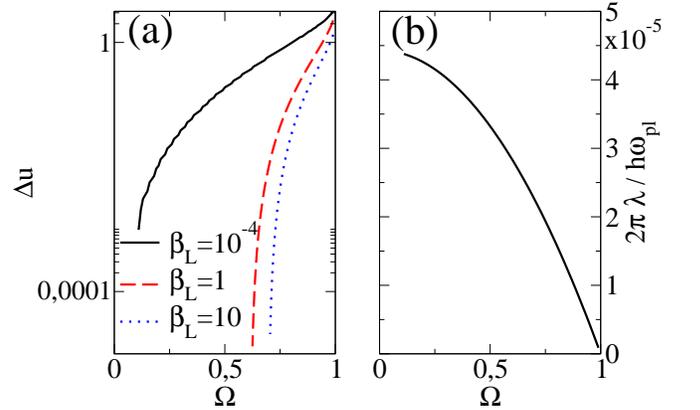}
 \caption{\label{fig:figure3} (a) the dimensionless depth
 $\Delta u$ of the minimum featuring maximum splitting $\lambda$ is
 plotted as a function of the ratio
 $\Omega=\omega/\omega_\mathrm{pl}$, for different values of the
 self-inductance parameter: $\beta_L=10^{-4}$ (solid line),
 $\beta_L=1$ (dashed line) and $\beta_L=10$ (dotted line). 
(b) the maximum value of the dimensionless splitting
 $\lambda/\hbar\omega_\mathrm{pl}$ is plotted as a function of
 $\Omega$; here $\beta_L=1$, but the result is independent of
 $\beta_L$.}  
\end{figure}

Fig.~\ref{fig:figure3}a  displays plots of the quantity
$\Delta u$, defined by Eq.~(\ref{eq:splitting}), as a function of
$\Omega$ for different values of the 
self-inductance parameter $\beta_L$; one sees that small values of
$\Omega$ correspond to very shallow minima in the potential energy,
whereas the deepest minima correspond to devices in which
$\omega=\omega_\mathrm{pl}$. However Fig.~\ref{fig:figure3}b shows
that for these devices no gap is 
expected in the first order in the coupling; this rather surprising
fact can be further illustrated by considering the dependence of the
matrix element $V'_{\gamma\xi}$ on $s=\sin\gamma\sin(\pi\phi)$ in
Eq.~(\ref{eq:Vp}): for $\Omega=1$, one has $\cos\gamma\cos(\pi\phi)=0$
and thus $s=0=\lambda$. Because
of this trade-off, the optimal condition corresponds to an intermediate
case. Note however that $\Delta u$ depends
on $E_J$ and $E_c$ only via the ratios
$\Omega=\omega/\omega_\mathrm{pl}$ and $I_b/I_c$ and thus can be tuned
independently of $E_J/E_c$. Therefore 
a bigger minimum depth can be obtained by
designing the SQUID so that the ratio $E_J/E_c$ is large. 

We now comment on the role of the self-inductance. 
The curve in Fig.~\ref{fig:figure3}b is very little
affected by the value of $\beta_L$, implying that this parameter is
not relevant for improving the splitting $\lambda$; on the contrary,
the curves in  Fig.~\ref{fig:figure3}a show that the dimensionless depth
$\Delta u$ of the minimum well is drastically reduced by increasing
the self-inductance of the loop. Thus loops of
smaller self-inductance are preferable.  

\begin{figure}[!tbh]
 \includegraphics[width=0.5\columnwidth]{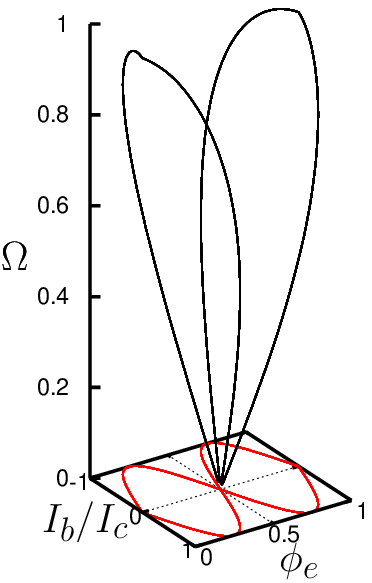}\nolinebreak
 \includegraphics[width=0.5\columnwidth]{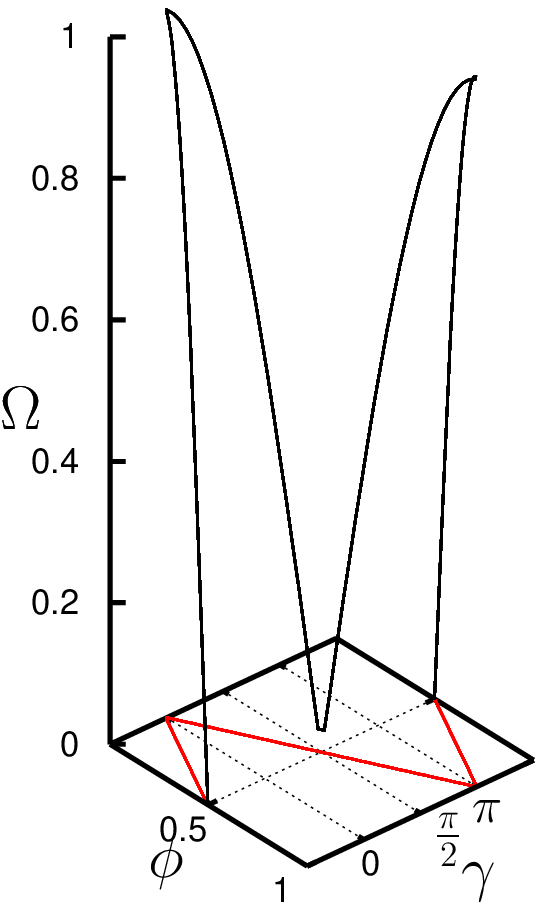}
 \caption{\label{fig:figure4} Plots of the solutions of
 Eqs.~(\ref{eq:eqs}) giving maximum value of the dimensionless
 splitting $\lambda/\hbar\omega_\mathrm{pl}$; left panel
 $(I_b/I_c,\phi_e,\Omega)$, right panel $(\phi,\gamma,\Omega)$. A
 projection of the points onto to the horizontal plane is also
 displayed for reader's convenience.}  
\end{figure}

Finally, Fig.~\ref{fig:figure4} represents numerical solutions of
Eqs.~(\ref{eq:eqs}) used in the previous figures. The
right panel shows the values of the dimensionless external parameters
$I_b/I_c$ and $\phi_e$ that corresponds to the maximum value of the
gap $\lambda$ for values of $\Omega$ between 0 and 1, while the
left panel shows a plot of the coordinates $(\bar\gamma,\bar\phi)$ of
the best minimum. As expected, the results are symmetric with respect
to simultaneous current inversion $I_b\to-I_b$ and flux reflection 
$\phi_e\to1-\phi_e$; this can be seen from
Eqs.~(\ref{eq:eqs}), where in this case if $(\gamma,\phi,\xi,I_b/I_c)$ is a
solution, then also $(\pi-\gamma,1-\phi,-\xi,-I_b/I_c)$ solves the
equations. 

In conclusion, we found that quantum oscillations of a NEMS embedded
into a SQUID can be detected by spectroscopic measurements in the
regime when one of the plasma frequencies of the SQUID matches the
frequency of the mechanical resonator. This frequency matching 
is possible with the current experimental techniques, and the scheme
has two parameters (external flux and external current) to
simultaneously tune the system to the vicinity of the degeneracy point
and perform spectroscopic measurements around this point. Measurements
of splitting of the degenerate doublet state displaying an avoided
crossing would be an unambiguous evidence of quantum nature of
mechanical vibrations. 

We acknowledge the financial support of the Future and Emerging
Technologies programme of the European Commission, under the FET-Open
project QNEMS (233992). We thank Herre van der Zant, Samir Etaki, and
Menno Poot for useful discussions.


\begin{thebibliography}{10}

\bibitem{Cleland}
A.~N. Cleland, \emph{Fundations of Nanomechanics}, Springer, Berlin (2003).

\bibitem{Blencowe04}
M.~Blencowe, Phys. Rep. \textbf{395}, 159  (2004).

\bibitem{Leggett02}
A.~J. Leggett, J. Phys: Condensed Matter \textbf{14}, R415 (2002).

\bibitem{Savel'ev06}
S.~Savel'ev, X.~Hu, and F.~Nori, New J. Phys. \textbf{8}, 105 (2006).

\bibitem{Cleland04}
A.~N. Cleland and M.~R. Geller, Phys. Rev. Lett. \textbf{93}, 070501 (2004);
M.~R. Geller and A.~N. Cleland, Phys. Rev. A \textbf{71}, 032311 (2005);

\bibitem{Pritchett}
E.~J. Pritchett and M.~R. Geller, Phys. Rev. A \textbf{72}, 010301 (2005).

\bibitem{Zou04}
X.~B. Zou and W.~Mathis, Phys.~Lett.~A \textbf{324}, 484 (2004).

\bibitem{Armour02}
A.~D. Armour, M.~P. Blencowe, and K.~C. Schwab, Phys. Rev. Lett. \textbf{88},
  148301 (2002).

\bibitem{Martin04}
I.~Martin {\em et al}, 
Phys.~Rev.~B \textbf{69}, 125339 (2004).

\bibitem{Rabl04}
P.~Rabl, A.~Shnirman, and P.~Zoller, Phys.~Rev.~B \textbf{70}, 205304 (2004).

\bibitem{Wei06}
L.~F. Wei {em et al}, 
Phys.~Rev.~Lett. \textbf{97}, 237201 (2006).

\bibitem{Armour08}
A.~D. Armour and M.~P. Blencowe, New J. Phys. \textbf{10}, 095004 (2008).

\bibitem{Hauss08}
J.~Hauss {\em et al}, 
New J. Phys. \textbf{10}, 095018 (2008).

\bibitem{Trees07}
B.~R. Trees {\em et al}, 
Phys. Rev. B. \textbf{76}, 224513 (2007); 
J.~Wabnig, J.~Rammer, and A.~L. Shelankov, {\em ibid} \textbf{75},
205319 (2007). 

\bibitem{Ruskov05}
R.~Ruskov, K.~Schwab, and A.~N. Korotkov, Phys.~Rev.~B \textbf{71}, 235407
  (2005).

\bibitem{Liao08}
J.~Q. Liao and L.~M. Kuang, Eur. Phys. J. B \textbf{63}, 79 (2008).

\bibitem{LaHaye09}
M.~D. LaHaye {\em et al},
Nature \textbf{459}, 960 (2009).

\bibitem{Zhou06}
X.~Zhou and A.~Mizel, Phys.~Rev.~Lett. \textbf{97}, 267201 (2006).

\bibitem{Buks06}
E.~Buks and M.~P. Blencowe, Phys.~Rev.~B \textbf{74}, 174504 (2006).

\bibitem{Xue07}
F.~Xue {\em et al},
Phys.~Rev.~B \textbf{76}, 064305 (2007).

\bibitem{Wang08}
Y.-D. Wang, K.~Semba, and H.~Yamaguchi, New J.~Phys. \textbf{10}, 043015
  (2008).

\bibitem{Zhang09}
J.~Zhang, Y.~X.~Liu, and F.~Nori, Phys. Rev. A \textbf{79}, 052102 (2009).

\bibitem{Pugnetti09b}
S.~Pugnetti {\em et al}, Phys. Rev. B \textbf{79}, 174516 (2009).

\bibitem{Etaki08}
S.~Etaki {\em et al},
Nature Phys. \textbf{4}, 785 (2008).

\bibitem{Mooij99}
J.~E. Mooij {\em et al},
Science \textbf{285}, 1036 (1999).

\bibitem{foot1} In principle one, many or no solutions can exist for a
  particular set of parameters; when more then one solution
  exist, we take the one with the bigger gap
  $\lambda$.

\end{thebibliography}
\end{document}